\documentclass[
 reprint,
 amsmath,amssymb,
 aps,superscriptaddress,
 citeautoscript,
twocolumn
]{revtex4-1}

\usepackage{graphics}
\usepackage{dcolumn}
\usepackage{bm}
\usepackage{amsmath}
\usepackage{epsfig}
\usepackage{braket}
\usepackage{commath}
\usepackage{color}
\usepackage{commath}
\usepackage{gensymb}
\usepackage{float} 

\usepackage[english]{babel}
\usepackage{color}

\usepackage{blindtext}

\usepackage{color}

\begin{document}

\title{High-harmonic generation enhancement with graphene heterostructures}

\author{I. Alonso Calafell}
\thanks{These authors contributed equally to this work.}
\affiliation{University of Vienna, Faculty of Physics, Vienna Center for Quantum Science and Technology (VCQ), 1090 Vienna, Austria}

\author{L. A. Rozema}
\thanks{These authors contributed equally to this work.}
\affiliation{University of Vienna, Faculty of Physics, Vienna Center for Quantum Science and Technology (VCQ), 1090 Vienna, Austria}

\author{A. Trenti}
\affiliation{University of Vienna, Faculty of Physics, Vienna Center for Quantum Science and Technology (VCQ), 1090 Vienna, Austria}
\affiliation{AIT Austrian Institute of Technology, Center for Digital Safety \& Security, Vienna, Austria}

\author{J. Bohn}
\affiliation{Department of Physics and Astronomy, University of Exeter, Exeter, EX4 4QL, UK}

\author{E. J. C. Dias}
\affiliation{ICFO-Institut de Ciencies Fotoniques, The Barcelona Institute of Science and Technology, 08860 Castelldefels (Barcelona), Spain}

\author{P. K. Jenke}
\affiliation{University of Vienna, Faculty of Physics, Vienna Center for Quantum Science and Technology (VCQ), 1090 Vienna, Austria}

\author{K.S. Menghrajani}
\affiliation{Department of Physics and Astronomy, University of Exeter, Exeter, EX4 4QL, UK}

\author{D. Alcaraz Iranzo}
\affiliation{ICFO-Institut de Ciencies Fotoniques, The Barcelona Institute of Science and Technology, 08860 Castelldefels (Barcelona), Spain}

\author{F. J. Garc\'{i}a de Abajo}
\affiliation{ICFO-Institut de Ciencies Fotoniques, The Barcelona Institute of Science and Technology, 08860 Castelldefels (Barcelona), Spain}
\affiliation{ICREA-Instituci\'o Catalana de Recerca i Estudis Avan\c{c}ats, Passeig Llu\'{\i}s Companys 23, 08010 Barcelona, Spain}

\author{F. H. L. Koppens}
\affiliation{ICFO-Institut de Ciencies Fotoniques, The Barcelona Institute of Science and Technology, 08860 Castelldefels (Barcelona), Spain}
\affiliation{ICREA-Instituci\'o Catalana de Recerca i Estudis Avan\c{c}ats, Passeig Llu\'{\i}s Companys 23, 08010 Barcelona, Spain}

\author{E. Hendry}
\affiliation{Department of Physics and Astronomy, University of Exeter, Exeter, EX4 4QL, UK}

\author{P. Walther}
\affiliation{University of Vienna, Faculty of Physics, Vienna Center for Quantum Science and Technology (VCQ), 1090 Vienna, Austria}

\date{\today}

\begin{abstract}
We investigate high-harmonic generation in graphene heterostructures consisting of metallic nanoribbons separated from a graphene sheet by either a few-nanometer layer of aluminum oxide or an atomic monolayer of hexagonal boron nitride. The nanoribbons amplify the near-field at the graphene layer relative to the externally applied pumping, thus allowing us to observe third- and fifth-harmonic generation in the carbon monolayer at modest pump powers in the mid-infrared. We study the dependence of the nonlinear signals on the ribbon width and spacer thickness, as well as pump power and polarization, and demonstrate enhancement factors relative to bare graphene reaching 1600 and 4100 for third- and fifth-harmonic generation, respectively. Our work supports the use of graphene heterostructures to selectively enhance specific nonlinear processes of interest, an essential capability for the design of nanoscale nonlinear devices.
\end{abstract}
\maketitle

High-harmonic generation (HHG) has been intensely investigated as a route towards the generation of coherent attosecond radiation in the extreme ultraviolet and x-ray spectral regions. Atomic gases have thus far been the most successful among demonstrated systems for HHG \cite{McPherson,Ferray_1988}, although they require high vacuum, making them impractical for the design of integrated devices. As a result, the development of solid-state HHG systems has become an important challenge, with promising results demonstrated for various crystalline materials \cite{ghimire2011observation,schubert2014sub,luu2015extreme}. However, the mechanisms of HHG in solid state appear to be fundamentally different from that in atomic gases, and are highly sensitive to crystal and polarisation orientations \cite{langer2016lightwave,you2017anisotropic}, as well as to the fundamental optical properties of the materials. In particular, theoretical models proposed for solid-state HHG have identified both interband transitions \cite{langer2016lightwave} and intraband electron dynamics \cite{you2017anisotropic} as radically different sources for the required anharmonicity.

Graphene constitutes an appealing choice of material in this context because it features sizeable intraband and interband contributions to the linear \cite{Jiang,Soavi,Mikhailov16,Rostami} and nonlinear \cite{mikhailov2008nonlinear,Marini17} conductivity in the infrared (IR). Moreover, as a two-dimensional material, it simplifies the conditions needed to achieve phase matching of the signal generated at different spatial locations (e.g., this is automatically guaranteed for pumping at normally incidence), which commonly limits the strength of nonlinear processes in thick crystals. Third- \cite{Jiang,Soavi,Kumar,Hong,Hendry,Baudisch,beckerleg2018cavity} and second-order \cite{Constant} nonlinearities have already been demonstrated in graphene, while higher-order harmonic generation has also been observed
\cite{bowlan2014ultrafast,hafez2018extremely,kovalev2021electrical,Baudisch,yoshikawa2017high}. These results have stimulated proposals for the exploitation of the nonlinear response of graphene to implement devices for quantum technology \cite{koppens, Gullans, Jablan, alonso2019quantum}.

The conical electronic band structure of graphene has been argued to boost the efficiency of intraband-mediated HHG \cite{mikhailov2008nonlinear, ishikawa2010nonlinear,AlNaib}. However, most measurements of HHG in graphene have identified a thermal origin and focused on THz frequencies \cite{bowlan2014ultrafast,hafez2018extremely,kovalev2021electrical}. Recent work has been reported on the observation of mid-IR HHG up to the fifth harmonic in multilayer graphene \cite{Baudisch}, as well as mid-IR HHG from monolayer graphene, though this has required extremely high peak intensities of $\sim 1\,$TW/cm$^2$ \cite{yoshikawa2017high}, at which ablation of the material is expected to take place \cite{roberts2011response}. An empirical demonstration of the sought-after efficient mid-IR HHG from monolayer graphene is still awaiting, to the best of our knowledge.

\begin{figure*}[t]
\centering
\includegraphics[width=\textwidth]{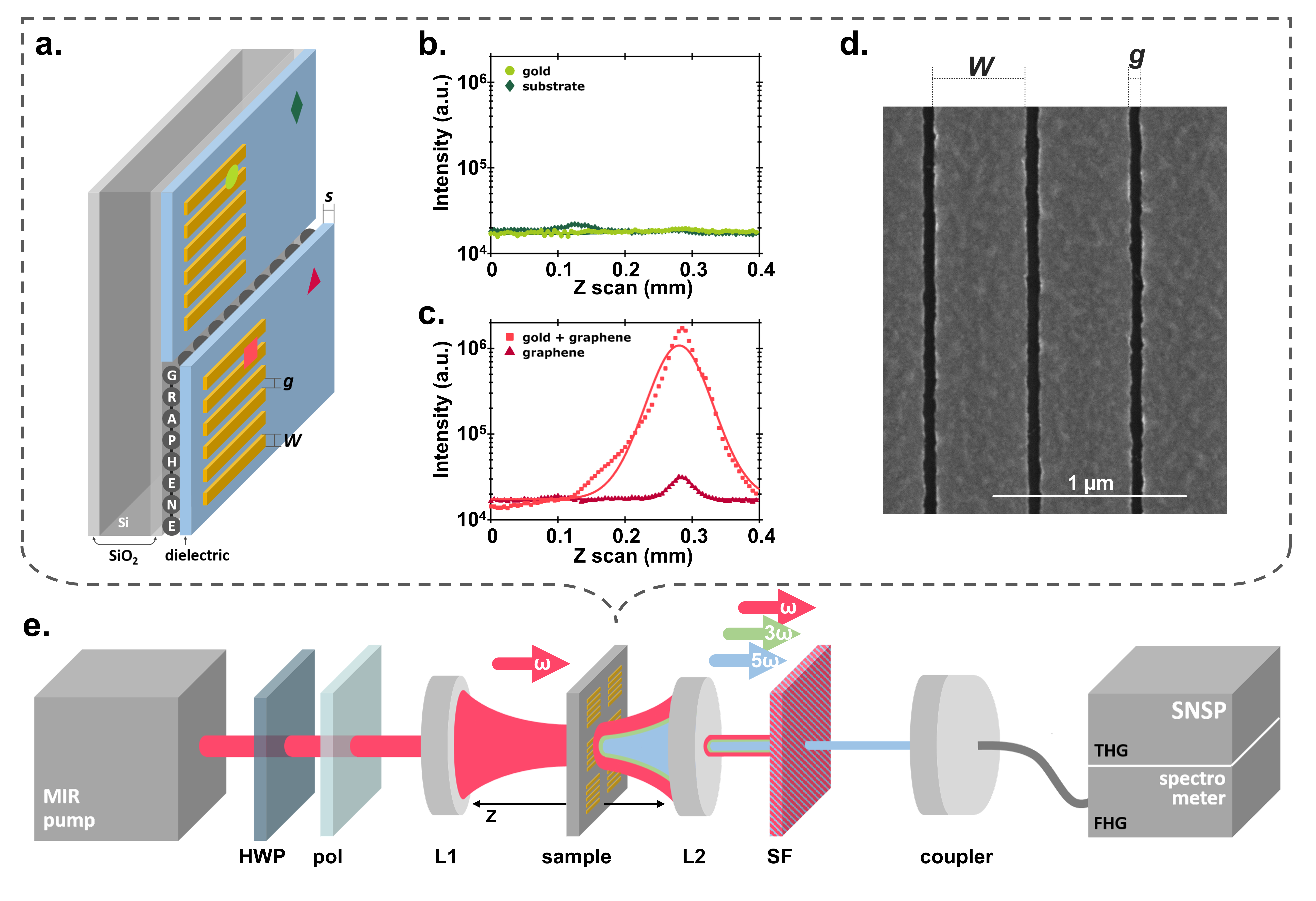}
\caption{\textbf{Graphene heterostructure sample and setup.} \textbf{a.} Graphene is separated by a dielectric material (h-BN or a $5\,$nm layer of $\mathrm{Al_{2}O_{3}}$) of thickness \textit{s} from an array of gold nanoribbons (width \textit{W}, gap \textit{g}) grown on top. Pump light is normally incident on the sample and the generated harmonics are collected in transmission.
\textbf{b.} The sample shows no third-harmonic signal when there is only substrate or gold nanoribbons. \textbf{c.} A significant third-harmonic signal is observed with bare graphene and a strongly enhanced signal when gold nanoribbons are added on top of the graphene. \textbf{d.} Scanning electron microscopy (SEM) image of an array. \textbf{e.} Experimental setup of the modified \textit{z}-scan measurement. An optical parametric oscillator provides 3900\,nm linearly-polarized femptosecond laser pulses at a 76\,MHz repetition rate. The half-wave plate (HWP), together with the polarizer (pol), allows rotating the polarization of the light that is then focused down with a short-focal lens (L1). A second lens (L2) collimates both the initial light and the THG signal excited in the sample that is moved in the \textit{z} direction, along the focal point of the beam. A spectral filter (SF) isolates the nonlinear signal that is then sent to either a superconducting nanowire single-photon detector (SNSPD) or a spectrometer to measure THG or FHG, respectively.}
\label{img:Figure1}
\end{figure*}

\begin{figure}[t]
\centering
\includegraphics[width=0.8\columnwidth]{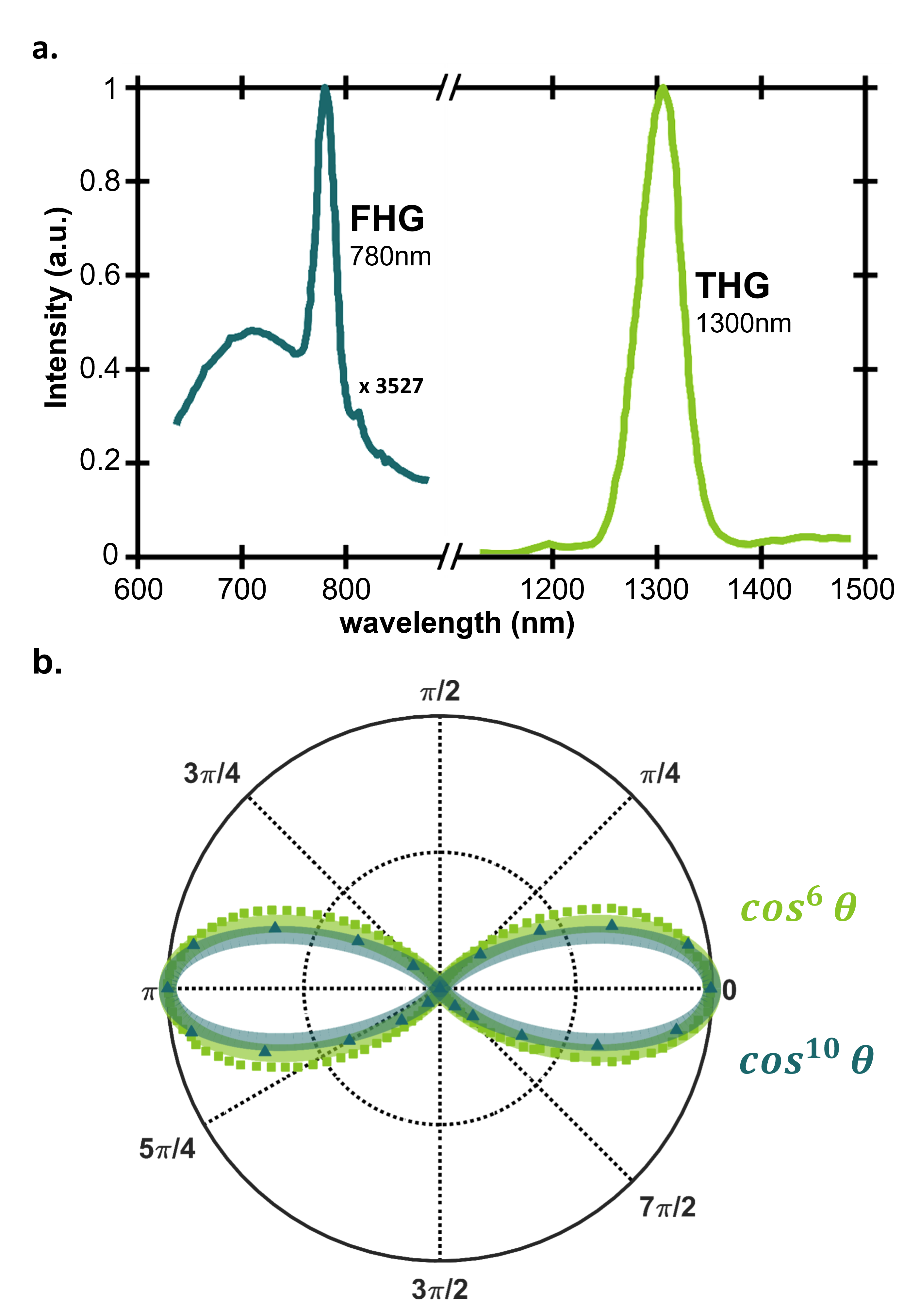}
\caption{\textbf{Characterization of the nonlinear harmonic signals.} \textbf{a.} Spectrum of the THG signal measured with a Fourier transform spectrometer ($15\,\mathrm{nm}$ resolution), along with the spectrum of the FHG signal measured with a single-photon spectrometer ($1\,\mathrm{nm}$ resolution). \textbf{b.} Typical THG and FHG signals as a function of input polarization relative to the ribbon array. As expected, the gold nanoribbons impose a $\cos^6{\theta}$ and $\cos^{10}{\theta}$ dependence for THG and FHG signals, respectively, maximizing the output when the incident light is perpendicular to the direction of the nanoribbons.
All of the data in this figure are taken with HBN\_70,
although the trends are the same with other parameters.}
\label{img:Figure2}
\end{figure}

\begin{figure}[t]
\centering
\includegraphics[
  width=0.9\columnwidth]{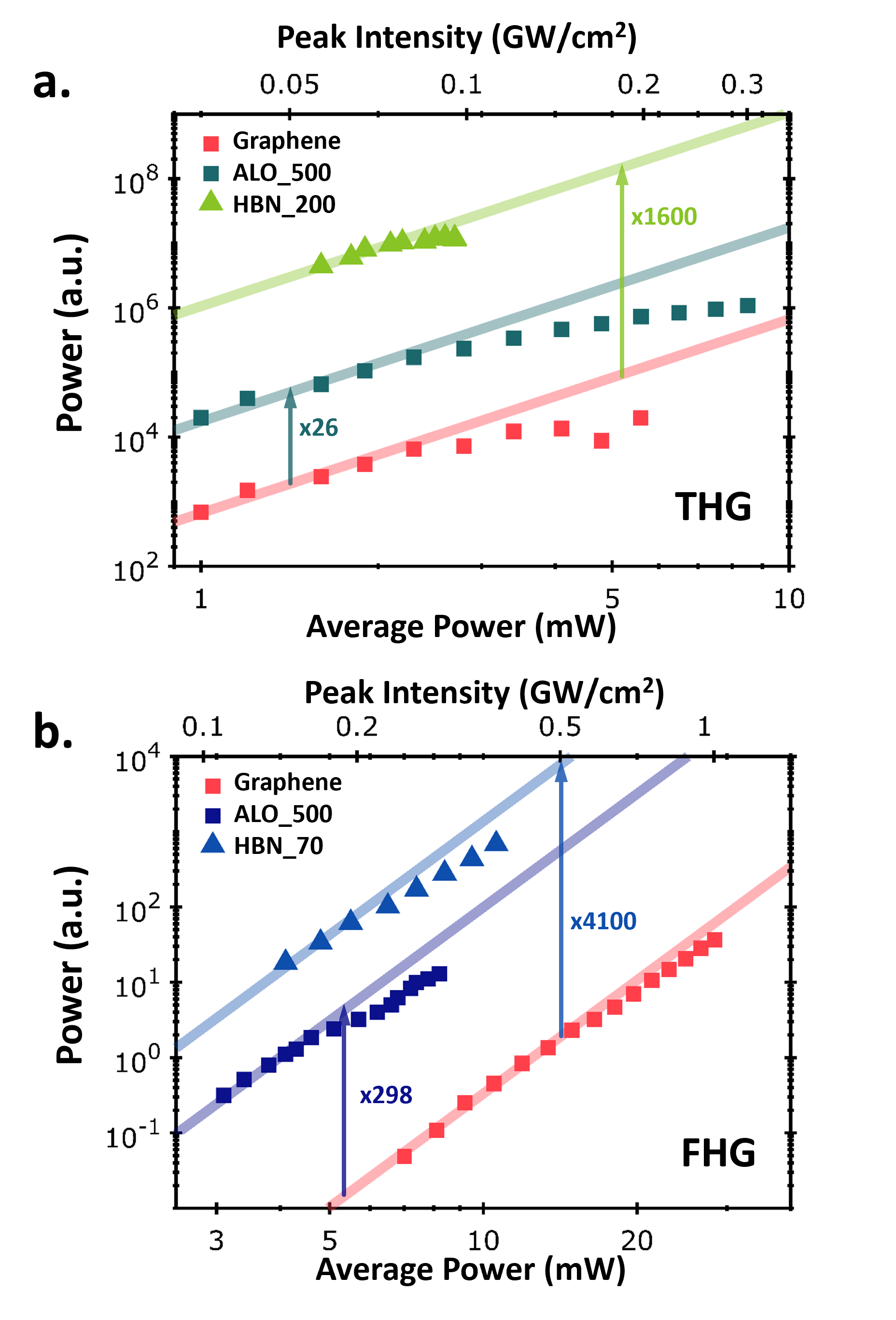}
\caption{\textbf{Spacer dependence of the nonlinear enhancement.} \textbf{a.} and \textbf{b.} show the THG and FHG measured power as a function of the input power for graphene with gold nanoribbons and bare graphene (red squares), respectively, for a $5\,\mathrm{nm}$ $\mathrm{Al_{2}O_{3}}$ spacer
(dark green/blue squares) and a monolayer h-BN spacer (bright green/blue triangles). 
For the THG measurements, we present the results from ALO\_500 and HBN\_200,
while for the FHG measurements, we show results from ALO\_500 and HBN\_70.
The pump wavelength is $3900\,$nm for all measurements.
In the device with $\mathrm{Al_{2}O_{3}}$ spacer, the THG and FHG signals are enhanced by a factor $26\pm1$ and $260\pm30$, respectively. For the h-BN spacer, the enhancement factors are $1600\pm500$ and $4100\pm600$, respectively. Error estimates are all $68\,$\% confidence intervals from the fit.}
\label{img:Figure3}
\end{figure}

\begin{figure*}[t]
\centering
\includegraphics[
  width=\textwidth]{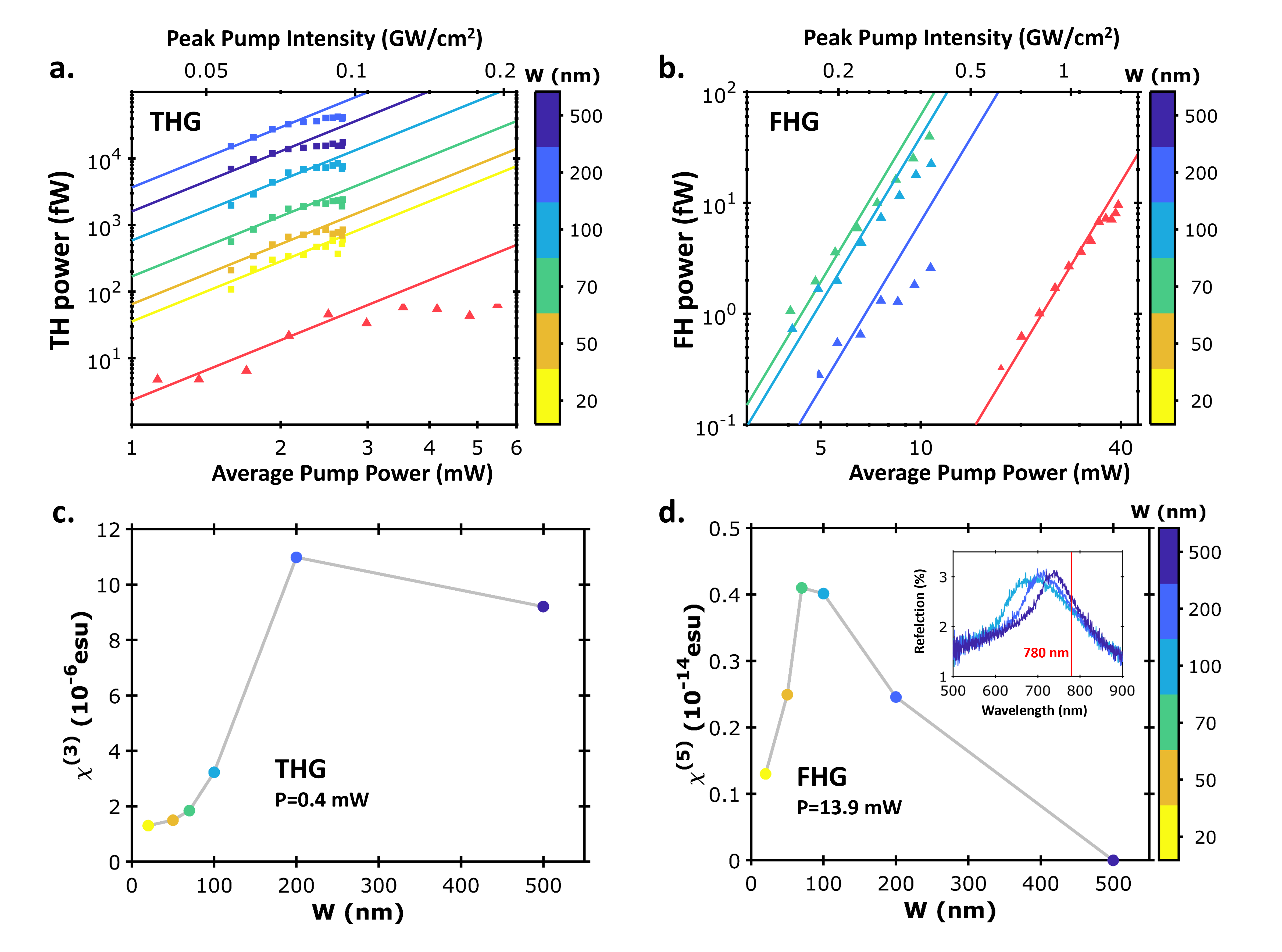}
\caption{\textbf{Nanoribbon width dependence of nonlinear enhancement with h-BN spacer.} \textbf{a.} and \textbf{b.} are the power dependence of the THG and FHG signals for bare graphene (red) and graphene with gold nanoribbons of different widths (yellow through blue).
For all the data in this figure, the pump wavelength is $3900\,$nm, resulting in third- and fifth-harmonic signals at $1300\,$nm and $780\,$nm, respectively.
The interception points found with the fits, whose slopes are set to 3 and 5, respectively, enable calculating the enhancement of the signals due to the field confinement. In THG, the maximum enhancement occurs at a nanoribbon width $W=200\,\mathrm{nm}$ and for FHG it occurs at $70\,\mathrm{nm}$. \textbf{c.} and \textbf{d.} summarize $\chi^{(3)}$ and $\chi^{(5)}$ for different nanoribbon widths and given powers $P=0.4\,\mathrm{mW}$ and $P=10.6\,\mathrm{mW}$, respectively.
\textbf{Inset:} Visible reflection spectra of the $100\,$nm, $200\,$nm, and $500\,$nm structures, colored light to dark blue as in the color bar.}
\label{img:Figure4}
\end{figure*}

In this paper, we report on the observation of fifth-harmonic generation (FHG) from monolayer graphene in the mid-IR spectral range with modest pumping peak intensities of $< 1\,$GW/cm$^2$. This result is made possible by the fabrication of heterostructures in which graphene is accompanied by metallic components to boost the external near field actually acting on the carbon sheet.
More precisely, we fabricate heterostructures made of graphene and gold nanoribbons that are separated by an insulator layer of either a $5\,$nm-thick layer of aluminium oxide {($\mathrm{Al_{2}O_{3}}$)} or a monolayer of {hexagonal} boron nitride {(h-BN)}. Similar heterostructures have also been used to launch acoustic graphene plasmons \cite{Alcaraz,lee2019graphene,epstein2020far}, observe plasmon-mediated third-harmonic generation (THG) \cite{alonso2021giant}, and enhance HHG at THz frequencies \cite{deinert2020grating}. Here, the nanoribbons act as mid-IR antennas, serving to enhance the incident pump field in the graphene layer and, consequently, boost the resulting harmonic generation. Importantly, we corroborate that ribbons exhibit no observable THG or FHG signals without the graphene layer. Upon investigation of the role of the insulator spacer thickness $s$ and the ribbon width $W$, we find that different ribbon widths are obtained to optimize the THG and FHG efficiencies, a surprising result that we attribute to an increased absorption of the fifth harmonic field by plasmonic modes in the metal ribbons at the generated optical wavelengths. Our findings are pivotal for the design of high-efficiency nanoscale nonlinear frequency-conversion devices.

A schematic of our heterostructure design is shown in Fig. \ref{img:Figure1}a. The samples consist of a $\mathrm{SiO_2}$-$\mathrm{Si}$-$\mathrm{SiO_2}$ substrate, on which a monolayer of {graphene is grown by chemical vapor deposition}. This is then covered by a dielectric: either a monolayer of h-BN or a $5\,$nm film of $\mathrm{Al_{2}O_{3}}$. On top of that, arrays of gold nanoribbons with different widths $W$ and fixed inter-ribbon gaps of $g=50\,$nm are etched using electron beam lithography and lift off (see Refs. \cite{Alcaraz,alonso2021giant} for more details). Fig. \ref{img:Figure1}b shows a typical scanning electron microscopy (SEM) image of the resulting high-quality metal edges of a representative sample.
We will refer to the different structures as follows: the samples with the $5\,$nm $\mathrm{Al_{2}O_{3}}$ spacer and ribbon width $W$ are called ALO\_W, while we refer to the samples with the monolayer h-BN spacer and ribbon width $W$ as HBN\_W.
For example, ALO\_200 is the heterostructure with a $5\,$nm $\mathrm{Al_{2}O_{3}}$ spacer, $g=50\,$nm and $W=200\,$nm.

To measure the nonlinear emission produced by our samples we use a modified $z$-scan setup, in which the nonlinear signal (either FHG or THG) is measured while the sample is moved along the \textit{z} axis through the focus of the laser beam (see Fig. \ref{img:Figure1}e). 
Our pump beam is a linearly-polarized pulsed laser with a $\sim 260\,$fs pulse duration, a central wavelength of $3900\,\mathrm{nm}$ ($0.225\,$eV), and a $76\,$MHz repetition rate.
This beam is generated by an optical parametric oscillator (OPO), fed by a mode-locked femtosecond Ti:saphhire laser. 
We use a half-wave plate (HWP) to rotate the polarization of the incoming beam to that set by the polarizer (pol).
A lens with a $5.95\,\mathrm{mm}$ focal length focuses the pump beam down to a waist of $\sim 13.5\,\mathrm{\mu m}$.
When the sample is moved parallel to the pump beam (along the $z$ axis), the nonlinear emission occurs most efficiently where the fluence is maximum (i.e. at the focal point). 
Afterwards, a lens with an $11\,\mathrm{mm}$ focal length collimates the beam, which is then sent through a set of spectral filters that separate the nonlinear emission from the pump beam. 
The resulting nonlinear signal is coupled into a multimode fiber, which can be sent to one of two detectors.
The THG signal (at $\sim 1300\,\mathrm{nm}$) is measured using a large-area SNSP detector from PhotonSpot with a $\sim 60\,\%$ detection efficiency at this wavelength.
We combine this detector with a Gemini interometer from NIREOS to measure the spectrum of the THG. 
To acquire the FHG signal, which is centered around $780\,\mathrm{nm}$, we connect the fiber to a single-photon sensitive silicon Andor spectrometer with a resolution of $1\,\mathrm{nm}$ and a detection efficiency of $\sim 10\,\%$.

To verify the origin of our nonlinear signals, we perform $z$-scans on various regions of our samples.
As shown in Fig. \ref{img:Figure1}b, neither the substrate (green diamonds) nor the gold nanoribbons without the graphene layer (green circles) display any measureable nonlinear response.  However, both graphene (red triangles) and graphene with nanoribbons (red squares) show a significant nonlinear signal at the focal point of the pump beam (see Fig. \ref{img:Figure1}c). 
These control measurements demonstrate that only the graphene layer contributes to any measurable nonlinearity in our setup. Moreover, they already show the very large nonlinear enhancement provided by the gold nanoribbons.

To further verify that our observed nonlinear signals are indeed associated with HHG, we measure their spectra. As displayed in Fig. \ref{img:Figure2}a, we find a THG signal at $\sim 1300\,\mathrm{nm}$ and a FHG signal at $\sim 780\,\mathrm{nm}$.
Both of them are at the expected wavelengths for our $3900\,\mathrm{nm}$ pump beam.
In addition to these two signals, we observe broadband white light generation due to thermal photoluminescence from the graphene layer. This has been previously reported in Ref. \cite{lui2010ultrafast}, and, as we discuss in the Appendix, the model presented therein fits our observed spectrum.

Owing to the geometry of the gold nanoribbons, the pump field enhancement is strongly dependent on the polarization of the input light. As shown in Fig. \ref{img:Figure2}b, the enhancement and the resulting nonlinear signal are maximized when the light polarization is perpendicular to the direction of the nanoribbons. However, when the polarization is rotated away from such perpendicular direction, the nonlinear signal decreases as $(\cos{\theta})^6$ and $(\cos{\theta})^{10}$, for the THG and FHG signals, respectively, reaching a minimum when the polarization is oriented along the nanoribbons. At this point, the signal strength is even lower than that of planar graphene without gold nanoribbons because of the screening produced by currents induced in the metal. 
In the Appendix, Fig. \ref{img:FigureS1} shows similar measurements performed on graphene without the nanoribbons, which demonstrate only a small (few percent) polarisation dependence of the nonlinear signals, attributed to birefringence of the silicon substrate.

To better understand the nature of the enhancement, we first study its dependence on the material and thickness $s$ of the dielectric spacer between the graphene and gold nanoribbons, and then proceed to characterize the dependence of the third- and fifth-harmonic signals on the nanoribbon width. To quantify the effect of the spacer material and thickness, in Fig. \ref{img:Figure3}a we show the enhancement of the THG signals for both ALO\_500 (dark green squares) and HBN\_200 (bright green triangles).
These are compared to the signal from bare graphene (red squares). We fit the nonlinear signals as a function of input power (straight lines) assuming a third order dependence and with the interception point in the logarithmic plot (which determines the magnitude of the third-order signal) as the only fitting parameter. The enhancement factor (EF) is then given by the difference between these interception points. We find an EF of $26 \pm 1$ with the $5\,\mathrm{nm}$ $\mathrm{Al_{2}O_{3}}$ spacer and an EF of $1600\pm500$ with the monlayer h-BN spacer.
In Fig. \ref{img:Figure3}b, we show a similar analysis for the FHG signals, assigning a fifth-order dependence of the signal. Here, we present measurements from ALO\_500  (dark blue squares)  and HBN\_70 (bright blue squares).
We find an EF of $260\pm30$ for the $5\,\mathrm{nm}$ $\mathrm{Al_{2}O_{3}}$ spacer and an EF of $4100\pm600$ with the h-BN spacer.
We discuss the width dependence in detail below.

Our power-scaling measurements indicate that we remain in the perturbative regime for both the THG and FHG measurements, in which the power of the $n^\mathrm{th}$-order harmonic scales with $p_\mathrm{pump}^n$, where $p_\mathrm{pump}$ is the average pump power \cite{wildenauer1987generation,li1989multiple}. Deviation from the expected power dependencies is only observed for the highest incident powers used here. As pointed out in Ref. \cite{alonso2021giant}, this is caused by an increased electron temperature \cite{yu2017ultrafast}, which reduces the nonlinearity. Efficient HHG for higher-harmonics in gas media typically takes place in the non-perturbative regime, wherein the power of all harmonics scales as $p_\mathrm{pump}^2$ \cite{higuchi2014strong}. This limit has also been reached in solid-state HHG \cite{schubert2014sub,liu2017high}, including in graphene \cite{yoshikawa2017high}. We were unable to reach this limit either in extended graphene or in our heterostructures without damaging the samples. In particular, we found that for, average pump powers above $\sim 30\,$mW ($\sim 1\,$GW/cm$^2$), the power of the THG and FHG signals began to permanently decrease after several seconds, presumably due to laser-induced structural modifications, as found in Ref.~\cite{roberts2011response}.

To quantify the effect of the nanoribbon width on the THG and FHG signal enhancement, in Fig. \ref{img:Figure4} we compile a series of measurements performed for devices with the monolayer h-BN spacer. 
For the $\mathrm{Al_{2}O_{3}}$ samples, we were only able to measure a FHG signal using ALO\_500.  In our attempts to measure FHG in the $\mathrm{Al_{2}O_{3}}$ samples with other widths, we observed laser-induced damage of the nanoribbons before detecting FHG.
As such, we will not discuss the width dependence in that sample.
In panels a and b, we show the third- and fifth-harmonic signals as a function of input power for nanoribbon widths of $20\,$nm to $500\,$nm with the monolayer h-BN spacer (yellow to blue) and bare graphene (red squares). Again, the points represent experimental data and the straight lines are fits, where only the interception points are free parameters (one per line) and the slopes are set to 3 and 5, respectively. Due to the long acquisition times and slow laser power fluctuations, we were unable to acquire reliable FHG power-scaling plots for 
HBN\_20, HBN\_50 and HBN\_500.
Note also that it was necessary to carry out these measurements in planar graphene using higher powers, which resulted in visible damage to these samples for the highest powers used. 

In panels c and d, we calculate the effective third-order and fifth-order susceptibilities of the graphene heterostructures \cite{Jiang,Boyd}, and find the maxima to be $\chi^{(3)}=5.6\cdot10^{-6}$ and $\chi^{(5)}=1.8\cdot10^{-14}\,\mathrm{esu}$. These values are several orders of magnitude larger than for bare graphene, highlighting the benefit of such nanostructures in enhancing the nonlinear properties of 2D materials.
Looking directly at the conversion efficiencies of the different processes (defined as the ratio of the generated TH or FH power over the input power), we find a maximum THG conversion efficiency of $\sim 2\times 10^{-8}$ with HBN\_200 at a pump power of $2.5$\,mW, and an FHG conversion efficiency of $\sim 4\times 10^{-12}$ in HBN\_70 with a pump power of $10$\,mW.
While these conversion efficiencies are lower than those that have been achieved in the THz regime \cite{bowlan2014ultrafast,hafez2018extremely,kovalev2021electrical}, they are significantly higher than in previous work within the mid-IR due to the enhancement produced by the heterostructures in our work.
For example, Ref. \cite{Baudisch} found THG and FHG conversion efficiencies of $3\times10^{-10}$ and $9\times10^{-14}$, respectively, in five-layer graphene. 

Interestingly, we observe that the optimal widths for harmonic generation is different for THG and FHG: the THG signal is strongest for ribbons with a width of $200\,\mathrm{nm}$, an effect which arises due to a maximum in field enhancement for larger ribbons \cite{alonso2021giant}. Somewhat unexpectedly, the FHG signal is strongest for the $70\,\mathrm{nm}$ ribbons. As the properties of pump beam are nominally identical (other than the power), this discrepancy most likely originates in the width-dependent additional absorption of the generated beams, as the metal nanoribbons in our heterostructures display plasmonic modes in the visible range. The width dependence of this plasmonic resonance is illustrated in the inset of Fig. \ref{img:Figure4}d, which shows the measured reflection spectra for samples with ribbons of widths $100\,$nm, $200\,$nm, and $500\,$nm. We observe that the peak of this plasmonic mode red shifts from about $675\,$nm for the $100\,$nm structure to $750\,$nm for the $500\,$nm structure.
The encroaching plasmon resonance attenuates the FHG signal (generated at $780\,$nm) more for the wider ribbons, while the THG signal does not undergo this additional absorption. This plasmon-quenching effect accounts for the weaker-than-expected FHG for wider ribbons.

In conclusion, we have demonstrated that graphene-based heterostructures can be used to enhance nonlinear conversion efficiencies in the mid-IR spectral range at relatively low pump powers. 
This has enabled us to observe FHG in the perturbative regime, wherein the generated nonlinear signal scales with the input pump power to the power of five. Such enhancement also allows us to observe third- and fifth-harmonic generation in graphene at modest pump powers in the mid-IR. We study the dependence of the nonlinear signals on the ribbon width and spacer thickness, as well as pump power and polarization, and demonstrate enhancement factors for third- and fifth-harmonic generation of up to 1600 and 4100, respectively, compared to bare graphene. 

We have shown that the ribbon width is one of the key parameters to adjust in order to optimise nonlinear conversion. For THG, we observe the largest nonlinear signals for ribbon widths of $\sim 200$ to $500\,$nm, an effect stemming from the optimal enhancement of the pumping field acting on the graphene. For FHG, we observe a maximum nonlinear conversion for smaller ribbon widths of $70\,$nm. For larger ribbon widths, the FHG is likely quenched due to spectral overlap with the plasmon resonance of the ribbons, which shifts towards the FHG emission wavelength of $780\,$nm for larger ribbons.  This demonstrates the fine tuning required in order to maximise a structure for a specific generation wavelength. This information is crucial in the design of future plasmonic, nanoscale HHG devices, particularly those intended to generate visible and near-IR wavelengths.

\section*{acknowledgments}
We acknowledge support from the Austrian Science Fund (FWF) through BeyondC (F7113) and Research Group 5 (FG5), the Air Force Office of Scientific Research (AFOSR) via PhoQuGraph (FA8655-20-1-7030),  the Engineering and Physical Sciences Research Council (EPSRC) of the United Kingdom via the EPSRC Centre for Doctoral Training in Metamaterials (Grant No. EP/L015331/1), the Spanish MICINN (PID2020-112625GB-I00 and SEV2015-0522), the Catalan CERCA, and Fundaci\'o Privada Cellex, the European Union’s Horizon 2020 research and  innovation program under the Marie Sklodowska‐Curie grant agreement No 801110 and the  Austrian Federal Ministry of Education, Science and Research (BMBWF).

\section*{Appendix}

\subsection{Extraction of the third-order and fifth-order susceptibilities---} Experimentally, $\chi^{(3)}$ and $\chi^{(5)}$ have been estimated following the procedure reported in Ref~\cite{Jiang}. The measured average power is proportional to the squared modulus of the electric field through the following relation:
\begin{equation}
P(\omega)=\frac{1}{8}\left(\frac{\pi}{\ln{2}}\right)^{3/2}f\tau W^2 n \epsilon_0 c \frac{|E(\omega)|^2}{2},
\label{eqn_PtoE}
\end{equation} 
where $f$ is the laser repetition rate, $\tau$ is the temporal pulse width, $W$ is the Gaussian beam diameter, $n$ is the refractive index, and $\epsilon_0$ and $c$ are the permittivity and speed of light in vacuum. In the experiment here reported, $f = 76\,\mathrm{MHz}$, $\tau = 260\,\mathrm{fs}$ (FWHM), $W = 27\,\mathrm{\mu m}$, and the refractive index $n = 2.4$ is considered to be constant.

The generated THG and FHG electric fields in a graphene monolayer of thickness $d_{gr} = 0.33\,\mathrm{nm}$ are related to the electric field at the fundamental pump frequency $E(\omega_0)$ through the following relations \cite{Boyd,Shen}:
\begin{subequations}
\label{eqn:chi3chi5}
\begin{align} 
E(3\omega_0) = i\frac{\omega_0}{8n c}\chi^{(3)} d_{gr} E(\omega_0)^3   \label{eqn:chi3}, \\
E(5\omega_0) = i\frac{\omega_0}{8n c}\chi^{(5)} d_{gr} E(\omega_0)^5  \label{eqn:chi5}.
\end{align}
\end{subequations}
From Eqs. (\ref{eqn:chi3chi5}), it is thus possible to calculate the effective $\chi^{(3)}$ and $\chi^{(5)}$ susceptibilities. Note that all the quantities in Eqs. (\ref{eqn_PtoE}) and (\ref{eqn:chi3chi5}) are given in SI units, while the $\chi^{(3)}$ and $\chi^{(5)}$ values in Fig. \ref{img:Figure3}(c-d) are given in the electrostatic system of units (esu). The expressions
\begin{subequations}
\label{eqn:chi35conv}
\begin{align} 
\chi^{(3)}(\mathrm{SI}) = \frac{4\pi}{(3*10^4)^2} \,\,  \chi^{(3)}(\mathrm{esu}),   \label{eqn:chi3conv} \\
\chi^{(5)}(\mathrm{SI}) = \frac{4\pi}{(3*10^4)^4} \,\, \chi^{(5)}(\mathrm{esu}), \label{eqn:chi5conv}
\end{align}
\end{subequations}
provide the relative conversion factors.

\begin{figure}[t]
\centering
\includegraphics[width=0.9\columnwidth]{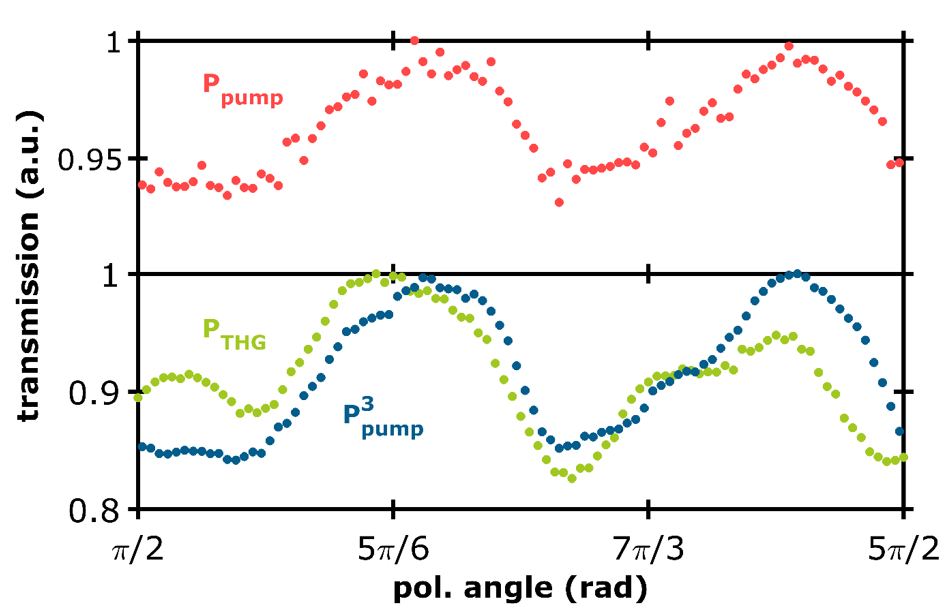}
\caption{\textbf{Polarization dependence of the transmission through the substrate.} Red points show the normalized polarization-dependent transmission through the substrate. Green points are the normalized THG power as a function of input polarization. Blue points are the cube of the transmitted pump power in the red points. All measurements are performed with a carrier wavelength of $3900\,\mathrm{nm}$.}
\label{img:FigureS1}
\end{figure}

\begin{figure}[t]
\centering
\includegraphics[width=0.9\columnwidth]{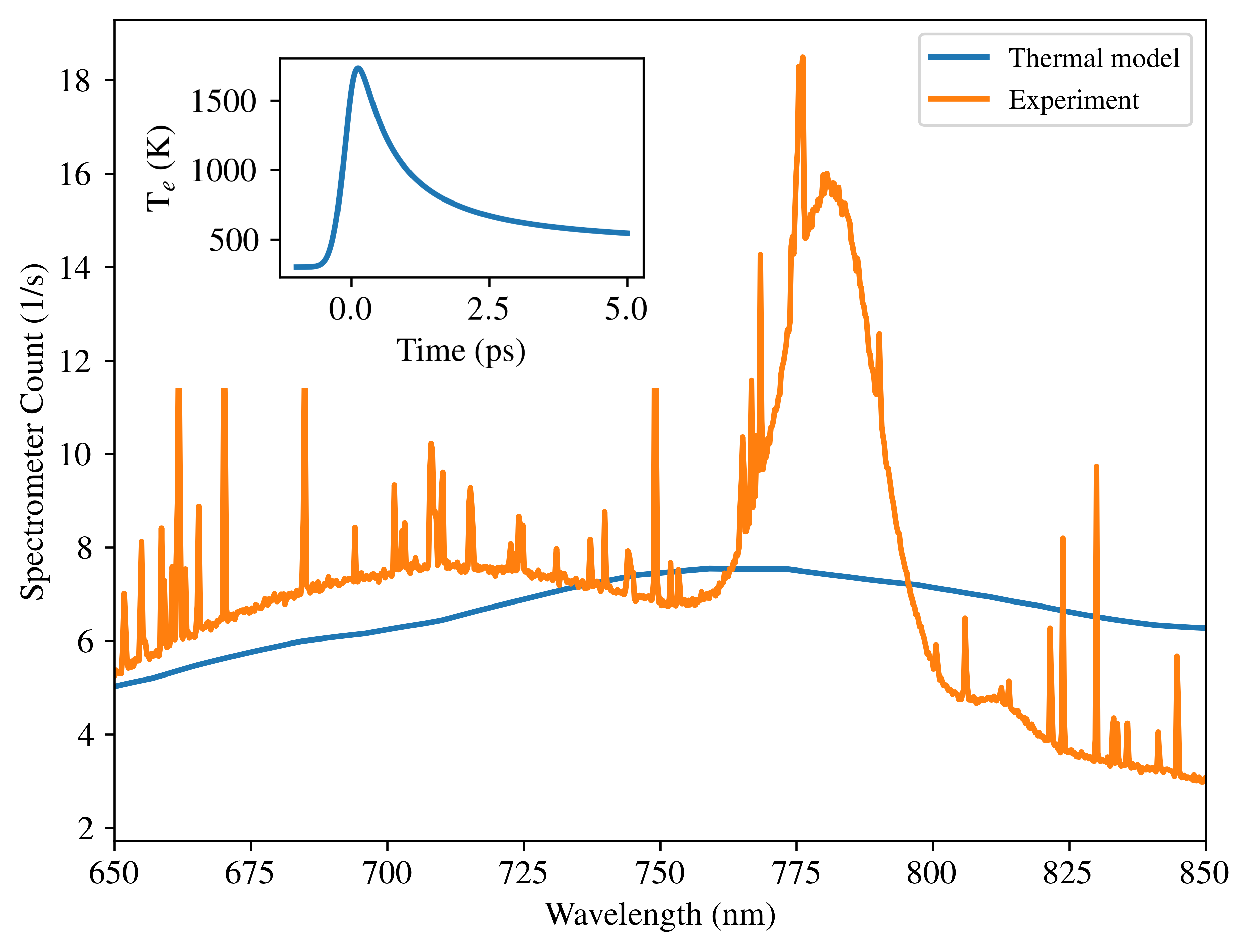}
\caption{\textbf{Ultrafast thermal emission from planar graphene.} Spectral emission of monolayer graphene on a Si wafer, pumped by $260\,$fs pulses with a peak intensity of $100\,$MW/cm$^2$. The experimental data (orange) shows not only the FHG signal around $780\,$nm, but also a broadband background. This latter is modelled as ultrafast thermal emission by hot carriers in graphene (blue). The inset shows the corresponding time-dependent electron temperature.}
\label{img:FigureS2}
\end{figure}

\subsection{Polarization dependence}

In this section, we discuss the polarization dependence of THG from bare graphene. The upper panel of Fig.~\ref{img:FigureS1} shows the transmission of the pump beam through the sample as a function of the orientation of its linear polarization. In this experiment, far-field light with a wavelength of $3900\,\mathrm{nm}$ is normally incident on the sample from the side of the substrate, composed of a stack of two layers of silica and one layer of silicon (see Fig. \ref{img:Figure1}). The silica layers have a thickness of $285\,\mathrm{nm}$, while the silicon layer has a thickness of $500\,\mathrm{\mu m}$ (see Ref. \cite{Alcaraz} for more detail of the sample fabrication). 
The measurements reported in Fig. \ref{img:FigureS1} are carried out on an area of the sample containing only graphene (i.e., there are no gold nanoribbons), just like in the violet triangles in Fig. \ref{img:Figure1}a of the main text. 
The red points in Fig.\ref{img:FigureS1} show a periodic pump transmission through the sample as a function of the incoming linear polarization angle. The measured relative difference in transmission is about 5\%. In the lower panel, the green points show the polarization-dependent normalized THG power together with the normalized cubed transmitted power (blue points). The quantitative agreement between the two explains the polarization dependence of the THG signal.

The physical origin of the polarization effect could be explained by a slight birefringence in the silicon Ref.~\cite{kruger}.
Another possible explanation is the dependence of the Fresnel reflection coefficient on polarization for off-normal incidence \cite{saleh}. 
Indeed, for an interface between air and silica, an incidence angle of $\sim 7\,\mathrm{\degree}$ relative to the interface normal is enough to induce a relative difference in reflectance by $5\,\mathrm{\%}$ between $s$ and $p$ polarizations.

\subsection{Thermal radiation background}

In this section, the physical origin of the background signal observed in the FHG spectrum of Fig. 2 is identified as ultrafast photoluminescence from graphene. The significant broadband light emission from graphene under excitation by femtosecond laser pulses has been studied in Ref. \cite{lui2010ultrafast} in great detail and is employed here to model our measurements. The ultrafast heating of the electron gas is modelled with a two-temperature model, comprising the electronic temperature $T_{el}$ and the temperature of strongly coupled optical phonons $T_{op}$:\\
\begin{eqnarray}
\frac{d T_{e l}(t)}{d t} &=&\frac{I(t)-\Gamma\left(T_{e l}, T_{o p}\right)}{c_{e}\left(T_{e l}\right)}, \\
\frac{d T_{o p}(t)}{d t} &=&\frac{\Gamma\left(T_{e l}, T_{o p}\right)}{c_{o p}\left(T_{o p}\right)}-\frac{T_{o p}(t)-T_{0}}{\tau_{o p}},
\end{eqnarray}
with a phonon decay time $\tau_{o p}=1\,$ps and a specific heat capacity of the phonons $c_{o p}\left(T_{o p}\right)$ and electrons $c_{e l}\left(T_{e l}\right)$.

We define the electron-phonon energy exchange rate $\Gamma$ as
\begin{widetext}
\begin{equation}
\begin{split}
	\Gamma\left(T_{e l}, T_{o p}\right) & =	\alpha\cdot (n_{o p}(T_{o p}) + 1) \int E\cdot(E - E_{o p})\cdot f_{e l}(E, T_{e l})\cdot [1 - f_{e l}((E - E_{o p}), T_{e l})]\text{d}E \\
	& - \alpha \cdot (n_{o p}(T_{o p})) \int E \cdot (E - E_{o p})\cdot f_{e l}(E - E_{o p}, T_{e l})\cdot [1 - f_{e l}((E), T_{e l})]dE,
\end{split}
\end{equation}
\end{widetext}
where $E_{o p} = 0.18$~eV is the phonon energy, $n\left(T_{o p}\right)=\left[\exp \left(E_{o p} / k T_{o p}\right)-1\right]^{-1}$ represents the phonon population at temperature $T_{o p}$, and $f\left(E, T_{e l}\right)=\left[\exp \left(E / k T_{e l}\right)+1\right]^{-1}$ is the Fermi-Dirac distribution for electrons at temperature $T_{el}$. The phonon coupling constant $\alpha$ corresponds to
\begin{equation}
\alpha = 3\dfrac{9 \left(\frac{\partial t}{\partial b}\right)^2}{\pi \rho E_{o p} \hbar^3 v_F^4},
\end{equation}
with $\frac{\partial t}{\partial b} = 45\,$eV/nm and $\rho = 7.6\times 10^{-7}~\text{kg}/\text{m}^2$ as the density of graphene [Ref2].

The calculated electron temperature in graphene for $260\,$fs pulses and an incident peak intensity of $100\,$MW/cm$^2$ is plotted in the inset of Fig. S\ref{img:FigureS2}. To get the corresponding spectral radiant fluency, we use Plancks's law:
\begin{equation}
\mathcal{F}\left(\lambda, T_{e l}\right)= \varepsilon(\lambda) \frac{2 h c^2}{\lambda^5}\int\left[\exp \left(\frac{hc/\lambda}{k T_{e l}(t)}\right)-1\right]^{-1}dt.
\end{equation}
The emissivity $\varepsilon(\lambda)$ of graphene is calculated via the absorption of light coming from air, taking into account the wavelength-dependent modulation introduced by the cavity formed by the $285\,$nm layer of glass between the graphene and Si. Assuming that the illuminated graphene area acts like a thermal Lambert emitter, we estimate our lens to collect up to an angle  $\theta = \arctan(\frac{D_{eff}/2}{f})$, where  $f = 11\,$mm is the focus length and the effective diameter is scaled to match the experimental signal via $D_{eff} = 30\% \times 5.5\,$mm. This is the only scaling parameter, which has been tweaked to match the thermal estimate with the background that we observe in the experiment, as shown in Fig. \ref{img:FigureS2}. It compensates alignment issues, as the apparatus is designed to measure HHG, but not white light generation. It also compensates for filter-reflection losses, coupling into the spectrometer, and similar effects. It does not compensate for wavelength-dependent performance, which we assume to be most influenced by the rough estimate of our absorbed mid-IR fluence ($\sim1.5\,$\%), but also by the collection lens having a chromatic alignment and wavelength-dependent efficiency. Taking all of these aspects into account, our estimate matches sufficiently well to identify ultrafast thermal emission of hot electrons as the dominant source of the background observed in the FHG measurements.

\bibliography{HHGpaper.bbl}

\end{document}